\documentclass[longauth]{aa} 

\usepackage{graphicx}
\usepackage{txfonts}

\newcommand{\sori}{S\,Ori~70}
\newcommand{\so}{$\sigma$~Orionis}
\newcommand{\mj}{$M_{\rm Jup}$}

\begin{document}

\title{New constraints on the membership of the T dwarf \sori~in the \so~cluster}

\subtitle{}

\author{M. R. Zapatero Osorio\inst{1}
       \and
        V. J. S. B\'ejar\inst{1}
       \and
        G. Bihain\inst{1,2}
       \and
        E. L. Mart\'\i n\inst{1,3}
       \and
        R. Rebolo\inst{1,2}
       \and
        I. Vill\'o-P\'erez\inst{4}
       \and
        A. D\'\i az-S\'anchez\inst{5}
       \and
        A. P\'erez Garrido\inst{5}
       \and
        J. A. Caballero\inst{6}
       \and
        T. Henning\inst{6}
       \and
        R. Mundt\inst{6}
       \and
        D. Barrado y Navascu\'es\inst{7}
       \and
        C. A. L. Bailer-Jones\inst{6}
       }

\offprints{M. R. Zapatero Osorio}

\institute{Instituto de Astrof\'\i sica de Canarias, E-38200 La Laguna, 
           Tenerife, Spain.\\
           \email{mosorio@iac.es}
      \and
           Consejo Superior de Investigaciones Cient\'\i ficas, Spain.
      \and
           University of Central Florida, Department of Physics, P.O. 
           Box 162385, Orlando, FL 32816, USA.
      \and
           Departamento de Electr\'onica, Universidad Polit\'ecnica de 
           Cartagena, E-30202 Cartagena, Spain.
      \and
           Departamento de F\'\i sica Aplicada, Universidad 
           Polit\'ecnica de Cartagena, E-30202 Cartagena, Spain.
      \and
           Max-Planck Institut f\"ur Astronomie, K\"onigstuhl 17, 
           D-69117 Heidelberg, Germany. Alexander von Humboldt Fellow.
           Now at Dept. Astrof\'\i sica, Univ. Complutense Madrid, Spain.
      \and
           LAEFF-INTA, P.O. 50727, 28080 Madrid, Spain
      }

\date{Received ; accepted }

\abstract
{}
{The nature of \sori~(S\,Ori~J053810.1$-$023626), a faint mid-T type
object found towards the direction of the young \so~cluster, is still
under debate. We intend to disentangle whether it is a field brown
dwarf or a 3-Myr old planetary-mass member of the cluster.}
{We report on near-infrared $JHK_s$ and mid-infrared [3.6] and
[4.5] IRAC/Spitzer photometry recently obtained for \sori. The
new near-infrared images (taken 3.82\,yr after the discovery
data) have allowed us to derive the first proper motion
measurement for this object. }
{The colors $(H-K_s)$, $(J-K_s)$ and $K_s$\,$-$\,[3.6] appear
discrepant when compared to T4--T7 dwarfs in the field. This
behavior could be ascribed to a low-gravity atmosphere or
alternatively to an atmosphere with a metallicity significantly
different than solar. The small proper motion of
\sori~(11.0\,$\pm$\,5.9~mas\,yr$^{-1}$) indicates that this
object is further away than expected if it were a single field T
dwarf lying in the foreground of the \so~cluster. Our measurement
is consistent with the proper motion of the cluster within
1.5\,$\sigma$ the astrometric uncertainty. }
{Taking into account \sori's proper motion and the new near- and
mid-infrared colors, a low-gravity atmosphere remains as the most
likely explanation to account for our observations. This supports
\sori's membership in \so, with an estimated mass in the interval
2--7~\mj, in agreement with our previous derivation.}

\keywords{stars: low mass, brown dwarfs -- stars: pre-main-sequence --
          open clusters and associations: individual ($\sigma$~Orionis) }

\maketitle

\section{Introduction}
Knowledge of the initial mass function is crucial for understanding
the formation processes of stars, brown dwarfs and free-floating
planetary-mass objects. Whether and where there is a limit for the
creation of objects by direct collapse and fragmentation of molecular
clouds has become one of the major goals in the study of very young
populations. Planetary-mass candidates with masses in the interval
3--13 Jovian masses (\mj) have been found in various star-forming
regions (e.g., Lucas \& Roche \cite{lucas00}; Zapatero Osorio et
al. \cite{osorio00}; Chauvin et al. \cite{chauvin04}; Lucas et
al. \cite{lucas05}; Luhman et al. \cite{luhman05}; Jayawardhana \&
Ivanov \cite{ray06}; Allers et al. \cite{allers06};
Gonz\'alez-Garc\'\i a et al. \cite{gonzalez06}; Caballero et
al. \cite{caballero07}). These objects are mostly free-floating but in
in a few cases appear as wide companions to young brown dwarfs or
low-mass stars.

\sori~(S\,Ori~J053810.1$-$023626) is the coolest free-floating,
planetary-mass candidate so far reported in the literature. It was
discovered by Zapatero Osorio et al$.$ (\cite{osorio02a}) and lies in
the direction of the \so~cluster (352~pc and 1--8~Myr, with a best
estimate at 3~Myr; Perryman et al. \cite{perryman97}; Oliveira et
al. \cite{oliveira02}; Zapatero Osorio et al. \cite{osorio02b}; Sherry
et al. \cite{sherry04}). The spectral type of \sori~was determined at
T5.5\,$\pm$\,1.0 from molecular indices measured over near-infrared
$H$- and $K$-band low-resolution spectra. Mart\'\i n \& Zapatero
Osorio (\cite{martin03}) obtained an intermediate-resolution spectrum
from 1.17 to 1.37~$\mu$m ($J$-band), in which the K\,{\sc i} doublet
at 1.25\,$\mu$m was detected. After comparison with theoretical
spectra from Allard et al. (\cite{allard01}), the authors inferred an
effective temperature and surface gravity of $T_{\rm
eff}$\,=\,1100\,$^{+200}_{-100}$~K and
log\,$g$\,=\,3.5\,$\pm$\,0.5~cm~s$^{-2}$, in agreement with the
expectations for a few megayears-old T dwarf. State-of-the-art
evolutionary models (Chabrier \& Baraffe \cite{chabrier00}; Burrows et
al. \cite{burrows97}; Baraffe et al. \cite{baraffe98}) yield a mass of
3\,$^{+5}_{-1}$~\mj~if \sori's very young age is finally confirmed.

Burgasser et al. (\cite{burgasser04}), in contrast, have raised
doubts about the low-gravity atmosphere and true cluster
membership of \sori. Based on the supposed similarity of the
observed spectra to field T6--T7 dwarfs, these authors argued
that the S\,Ori object is ``an old, massive field brown dwarf
lying in the foreground of the \so~cluster''. However, this work
relied on low signal-to-noise ratio data. Better quality
photometry and spectra are needed to assess the true nature of
this candidate.

Here we present astrometric measurements, IRAC/Spitzer data and
$JHK_s$ photometry for \sori. We find that this object has colors
unexpected for its spectral classification, which is measured in the
range T4.5--T7 with a best estimate at T6. We ascribe this to a low
gravity atmosphere, with a different metallicity being an alternative,
but less likely, explanation.

\begin{table*}
\caption[]{Log of near-infrared observations of \sori.}
\label{obslog}
\centering
\begin{tabular}{llcccc}
\hline\hline
Telescope   & Instrument & Field of view & Pixel    & Observing dates & Exposure time \\
            &            & (arcmin$^2$)  & (arcsec) &                 & (s)           \\
\hline                  
3.5\,m CAHA & Omega-2000 & 225           & 0.45     & 2005 Oct 22 ($JH$), 2005 Oct 25 ($K_s$) & 3600 ($J$), 6000 ($H$), 7200 ($K_s$) \\
10\,m KeckII& NIRSPEC    & 0.59          & 0.18     & 2005 Oct 26 ($JK'$)                     & 405 ($J$), 270 ($K'$)                \\
\hline                  
\end{tabular}
\end{table*}

\begin{table*}
\caption[]{Near-infrared (2MASS photometric system) and IRAC/Spitzer photometry of \sori.}
\label{phot}
\centering
\begin{tabular}{lccccccc}
\hline\hline
Telescope  &         $J$        &         $H$        &        $K_s$       &         [3.6]       &        [4.5]       &         $H-K_s$      &         $J-K_s$      \\
           &        (mag)       &        (mag)       &        (mag)       &         (mag)       &        (mag)       &          (mag)       &          (mag)       \\
\hline                    
3.5 m CAHA & 19.98\,$\pm$\,0.06 & 20.07\,$\pm$\,0.07 & 19.60\,$\pm$\,0.08 &      ...            &       ...          & $+$0.48\,$\pm$\,0.11 & $+$0.38\,$\pm$\,0.10 \\
Keck\,II   & 19.96\,$\pm$\,0.07 &       ...          & 19.58\,$\pm$\,0.07 &      ...            &       ...          &        ...           & $+$0.38\,$\pm$\,0.10 \\
Spitzer    &        ...         &       ...          &        ...         &  18.62\,$\pm$\,0.30 & 17.19\,$\pm$\,0.15 &        ...           &         ...          \\
\hline                    
\end{tabular}
\end{table*}

\section{Observations}
\subsection{Near-infrared photometry}
\sori~was observed in $J$, $H$, and $K_s$ broad-band filters with
prime focus wide field camera Omega-2000 (2048$\times$2048 pixels;
Bailer-Jones et al. \cite{bailer00}) on the 3.5~m telescope at the
Calar Alto (CAHA) Observatory. We also imaged \sori~with the NIRSPEC-3
(similar to $J$, see Fig.~\ref{filters}) and $K'$ filters and the
slit-viewing camera (256$\times$256 pixels) of the near-infrared
spectrometer NIRSPEC (McLean et al. \cite{mclean00}) on the Keck~II
telescope (Hawai'i). The observing log containing instrumental
information, dates of observations, and exposure times per filter is
provided in Table~\ref{obslog}. Images ($J$-band) of the T4.5 spectral
standard dwarf 2MASS\,J05591914$-$1404488 (J0559$-$14 from now on)
were collected with NIRSPEC immediately after \sori; total exposure
time was 27~s. During both the CAHA and Keck observations the weather
was photometric and the average seeing was 0\farcs8 (CAHA) and
0\farcs5 (Keck) in the $K$-band. Raw data were reduced in a standard
fashion and included both sky subtraction, flat-fielding, and
combination of individual frames.

We performed point-spread-function (PSF) photometry using
{\sc iraf}\footnote{IRAF is distributed by National Optical Astronomy
Observatories, which is operated by the Association of Universities
for Research in Astronomy, Inc., under contract to the National
Science Foundation, USA}. CAHA and Keck instrumental magnitudes were
transformed into observed magnitudes using between 1 and 33 2MASS
sources (depending on the filter) that were present within the fields
of view. The dispersion of the photometric zero point is $\pm$0.05~mag
in all bands. T dwarfs exhibit complex spectral energy distributions
within the near-infrared. To account for this effect and in order to
transform the CAHA and Keck magnitudes into the 2MASS photometric
system, we applied the following correction factors: $-$0.04~mag
($J$), $+$0.02~mag ($H$), and $+$0.11~mag ($K_s$) for CAHA data,
$+$0.02~mag ($J$) and $-$0.03~mag ($K'$) for Keck data. These were
obtained via the integration of observed near-infrared spectra of both
main-sequence stars of different spectral types as well as T dwarfs
(Pickles \cite{pickles98}; Leggett et al. \cite{leggett02}; Geballe et
al. \cite{geballe02}), convolved with the Omega-2000, NIRSPEC and
2MASS filter passbands and atmospheric transmission curves of the
observatories. All filter passbands are plotted in
Fig.~\ref{filters}. Corrections are expected to be larger for the $J$
and $K$ filters than for the $H$-band. Nevertheless, we note that
color corrections are similar or slightly smaller than the measured
photometric uncertainties. CAHA and Keck magnitudes (2MASS photometric
system) of \sori~are reported in Table~\ref{phot}. As a test of
consistency, we also calibrated the Keck $J$ magnitude of \sori~using
the observations of J0559$-$14 (T4.5, any color effect should be
minimal), deriving $J$\,=\,19.89\,$\pm$\,0.05~mag, which coincides
within the observed errors with the values given in
Table~\ref{phot}. The excellent agreement in $J$ and $K$ between the
CAHA and Keck data indicates that no strong sistematic errors are
affecting the measurements. Our data suggests that \sori~does not show
significant photometric variability over a few days ($\le$0.1~mag).

\subsection{Mid-infrared photometry}
We derived mid-infrared photometry using public mosaic images
available from the Spitzer Space Telescope Data Archive, taken on 2004
October 9 under the Spitzer Guaranteed Time Observation program
\#37. Hern\'andez et al. (\cite{hernandez07}) provides information on
data aquisition, exposure times, and data reduction. We downloaded
mosaic images in the four channels (3.6, 4.5, 5.8, and 8.0~$\mu$m) of
the IRAC instrument (Fazio et al. \cite{fazio04}) using the Leopard
software. \sori~is only detected in the [3.6] and [4.5] bands with
signal-to-noise ratios of 9 and 16 in the peak flux. We performed
aperture photometry on various single, bright sources near \sori~using
$daophot$ in {\sc iraf}, with an aperture radius of 12\arcsec~and a
background annulus extending from 12 to 24\arcsec.  For the [3.6] and
[4.5] bands we adopted {\sl daophot} zero point magnitudes of 17.30
and 16.82, respectively, as described at
http://ssc.spitzer.caltech.edu/archanaly/quick.phot. We derived the
PSF from the brighter sources and applied this to determine PSF
photometry for \sori. Results are reported in Table~\ref{phot}.

\subsection{Proper motion}
We determined the proper motion of \sori~via the comparison of its
position with respect to four common point-like sources in various
near-infrared images obtained at different epochs. As a starting point
for the astrometric analysis, we used the $K_s$-band frame shown in
Fig.~1 of Zapatero Osorio et al. (\cite{osorio02a}), which was
obtained on 2001 December 29 with the Keck I telescope and has an
average seeing of 0\farcs8 and a field of view of
32\arcsec$\times$32\arcsec. The angular resolution was
0\farcs15~pix$^{-1}$. The three recent CAHA $JHK_s$ images and the
Keck $J$ and $K'$ frames reported here act as the second epoch
astrometric data. Hence, the time interval spanned by the available
observations is 3.82~yr. Our proper motion measurements are shown in
Table~\ref{pm} using the CAHA and Keck data separately as well as the
combined result.  The associated uncertainty is derived from the
dispersion of the four reference stars around null motion and the
dispersion of the proper motion measurements obtained for \sori~using
the various images ($JHK_sK'$) available per epoch.

\begin{table}
\caption[]{Proper motion of \sori.}
\label{pm}
\centering
\begin{tabular}{lcc}
\hline\hline
Telescope  & $\mu_{\alpha} \rm{cos}\delta $ & $\mu_\delta$  \\
           & (mas~yr$^{-1}$)               & (mas~yr$^{-1}$)\\
\hline
3.5 m CAHA & $+$9.9\,$\pm$\,2.6            & $+$6.0\,$\pm$\,5.6 \\
Keck\,II   & $+$9.8\,$\pm$\,4.2            & $+$3.7\,$\pm$\,4.2 \\
Average    & $+$9.8\,$\pm$\,3.4            & $+$5.0\,$\pm$\,4.9 \\
\hline
\end{tabular}
\end{table}

\begin{figure*}
\centering
\includegraphics[width=5.9cm]{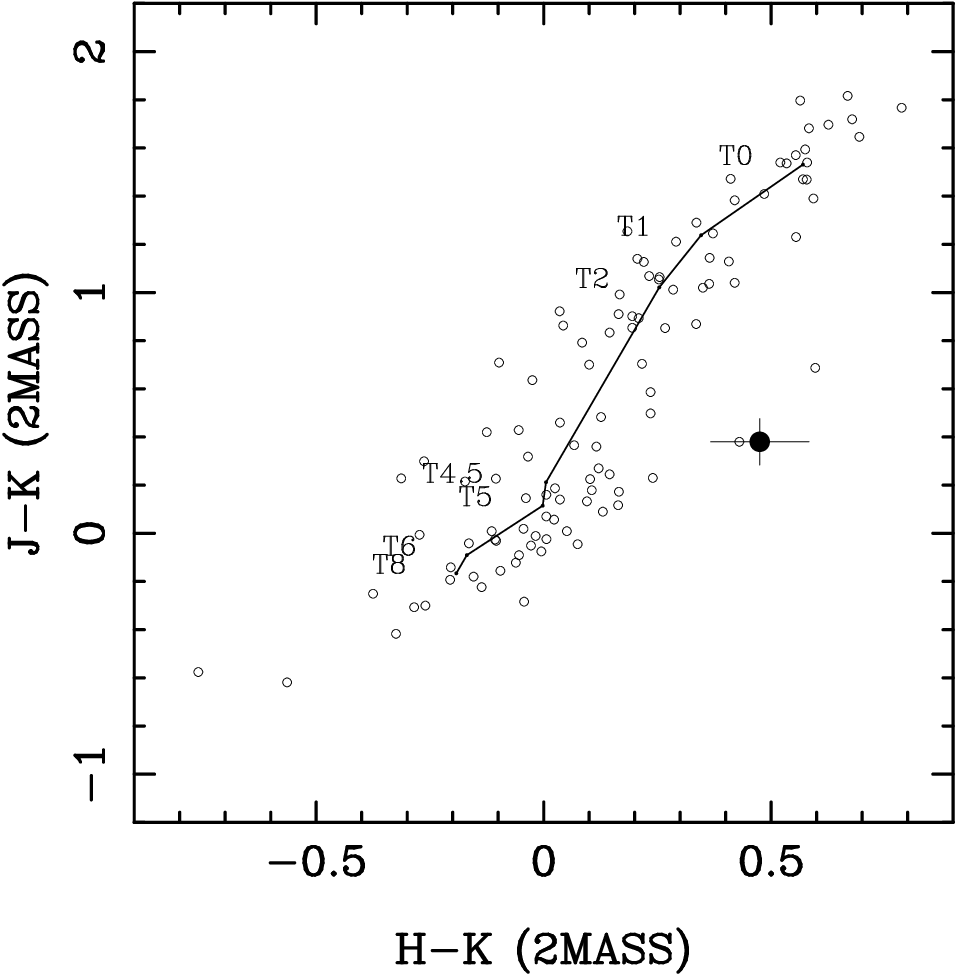}~~
\includegraphics[width=5.9cm]{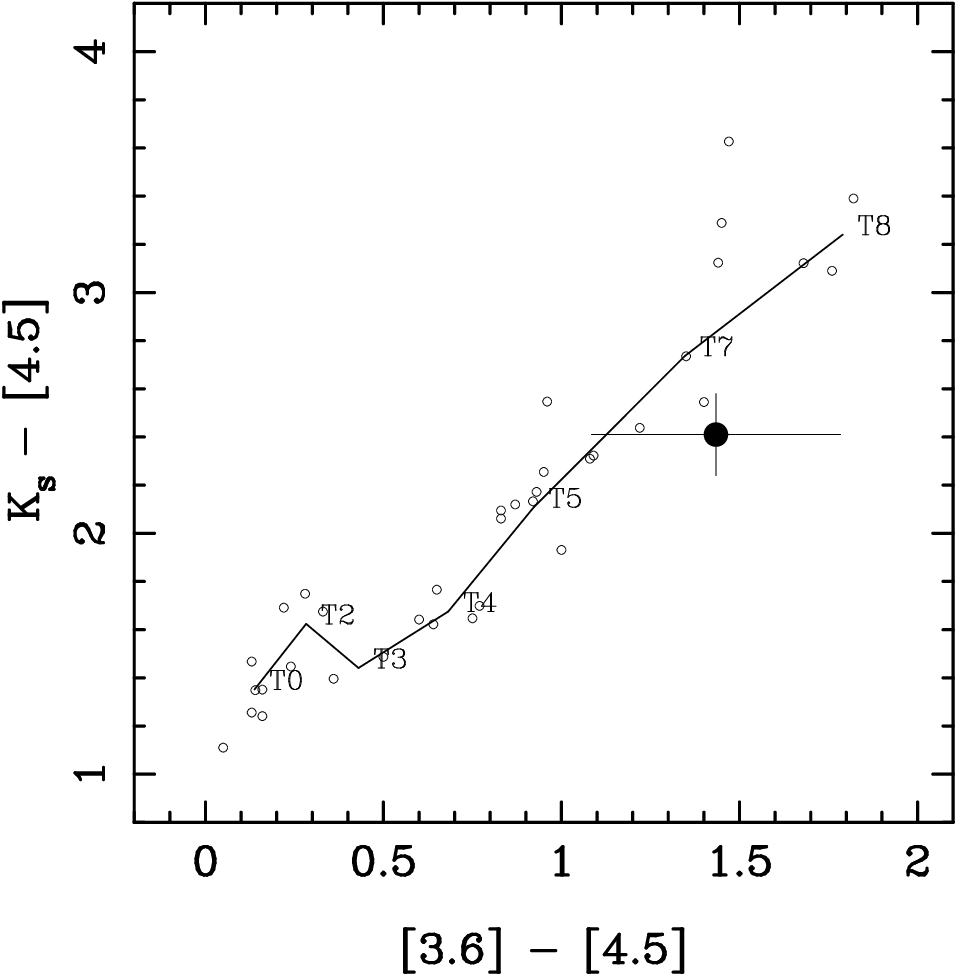}~~
\includegraphics[width=5.9cm]{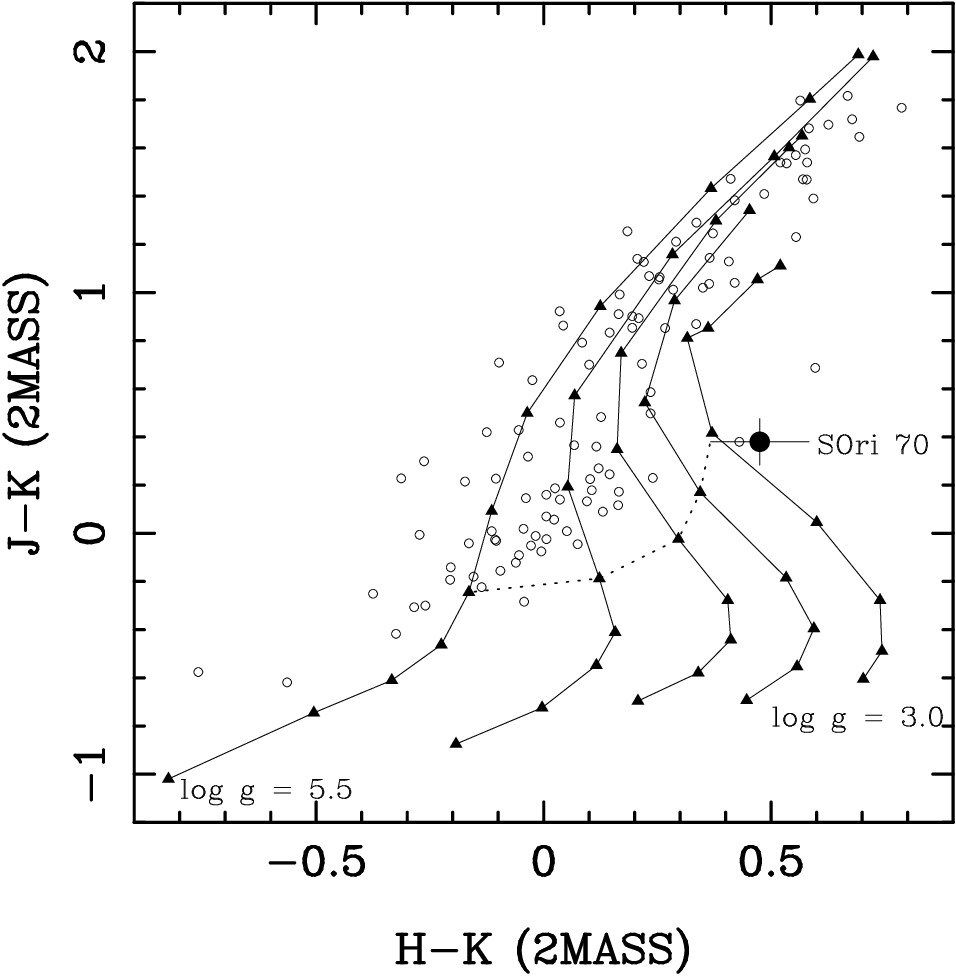}
\caption{Color-color diagrams of \sori~(solid circle) and field
T-dwarfs (open circles). For clarity, only the error bars of \sori~are
plotted. In the left and middle panels, the sequence defined by the
field brown dwarfs is displayed with a solid line. Spectral types are
indicated. The right panel shows the solar-metallicity models by Tsuji
et al. (\cite{tsuji04}) superimposed onto the data. Gravities of
log\,$g$\,=\,3.0, 3.5, 4.0, 4.5, and 5.5 cm\,s$^{-2}$ are depicted
with solid lines. The triangles along each model curve are in $T_{\rm
eff}$ invervals of 100\,K, ranging from 700 (bottom) to 1500\,K
(log\,$g$\,=\,3.0), 1400\,K (log\,$g$\,=\,3.5), 1600\,K
(log\,$g$\,=\,4.0 and 4.5), and 1700\,K (log\,$g$\,=\,5.5). The
log\,$g$\,=\,5.0 track is not plotted for clarity. The dotted line
connects the $T_{\rm eff}$\,=\,1100\,K points in each model.}
\label{colcol}
\end{figure*}

\section{Discussion}
The low proper motion of
\sori~($\mu$\,=\,11.0\,$\pm$\,5.9~mas~yr$^{-1}$) makes it unlikely
that it is a nearby ($\le$30~pc) T dwarf. On the one hand, we compared
our measurement with the motion of 192 Hipparcos stars (Perryman et
al. \cite{perryman97}) within a radius of $15^\circ$ around \so~and at
a distance between 80 and 130~pc, which is the distance interval
expected for \sori~if it were a field, single T6 dwarf. About
70\%~of the Hipparcos stars show larger motion than \sori, suggesting
that the S\,Ori object is located farther away. On the other hand, our
measurement is consistent (within 1.5~$\sigma$) with the proper motion
of the O9.5V--B0.5V-type star \so~AB, which is the most massive member
of the cluster of the same name. We note, however, that the relative
motion of the Orion OB association is directed away from the Sun (de
Zeeuw et al. \cite{zeeuw99}). This makes it very hard to detect
cluster members via proper motion analysis. On the contrary, radial
velocity studies can be more discriminant (Jeffries et
al. \cite{jeffries06}) but the extreme faintness of \sori~prevents any
accurate radial velocity measurement with current instrumentation.

\subsection{Color-color diagrams}

Color-color diagrams are depicted in Fig.~\ref{colcol}. To put
\sori~into context, we included $JHK_s$ data of more than 100 field
T-dwarfs and IRAC/Spitzer photometry of 36 field T-dwarfs compiled
from the literature (Knapp et al. \cite{knapp04}; Tinney et
al. \cite{tinney05}; Patten et al. \cite{patten06}; Burgasser et
al. \cite{burgasser06a}; Artigau et al. \cite{artigau06}; Mugrauer et
al. \cite{mugrauer06}; Leggett et al. \cite{leggett99},
\cite{leggett02}, \cite{leggett07}; Liu et al. \cite{liu07}; Luhman et
al. \cite{luhman07}; Looper et al. \cite{looper07}). All near-infrared
colors were conveniently transformed into the 2MASS photometric system
using equations quoted in Stephens \& Leggett (\cite{stephens04}),
which are appropriate for ultracool dwarfs.

The location of \sori~in Fig.~\ref{colcol} is challenging since this
object appears as an outlier, particularly when the $K$-band magnitude
is involved. Two field dwarfs lie near it in the near-infrared
color-color panels: 2MASS\,J00501994$-$3322402 (Tinney et
al. \cite{tinney05}), whose photometric errors are quite large, and
2MASS\,J13243559$+$6358284 (Looper et al. \cite{looper07}). The latter
object is widely discussed by Looper et al. (\cite{looper07}) in terms
of binarity and low-gravity atmosphere. We applied the criterion
defined by Covey et al. (\cite{covey07}, eq.~2) to distinguish objects
with photometric properties deviating from the properties typical of
field dwarfs and found that \sori~lies more than 2\,$\sigma$ away from
the near-infrared sequence defined by the field T-type brown
dwarfs. In the mid-infrared wavelengths, the photometry of
\sori~deviates from the field on a 1--2\,$\sigma$ level. Covey et
al.'s equation takes into account the color uncertainties of \sori~and
the width of the field distribution.

As compared to T4--T7 field dwarfs, \sori~presents redder $(H-K_s)$,
$(J-K_s)$, and [3.6]\,$-$\,[4.5] colors and a bluer $K_s$\,$-$\,[3.6]
index than expected for its spectral type ($\sim$T6). However,
the $(J-H)$ index is similar to that of T5--T6 field dwarfs. The
$K$-band reddish nature of \sori~is also apparent in its
low-resolution $HK$ spectrum. Fig.~4 of Burgasser et
al. (\cite{burgasser04}) shows these data along with the spectrum of
the field T6.5 2MASS\,J10475385$+$2124234. Both spectra were obtained
with similar instrumentation and are normalized to unity at
1.57~$\mu$m. While the field dwarf matches reasonably well the
$H$-band region of \sori, it underestimates the flux at $K$-band,
supporting the redder $(H-K_s)$ index of \sori. Burgasser et
al. (\cite{burgasser04}) argued that this may be indicative of a
``lower surface gravity for \sori~relative to
2MASS\,J10475385$+$2124234''. 

Multiplicity cannot explain the observed photometric properties of
\sori. We artificially produced near-infrared colors of L--T and T--T
pairs using the absolute magnitudes provided by Liu et
al. (\cite{liu06}). None of the combinations were able to reproduce
our observations.

The comparison of our data to theory is shown in the right panel of
Fig.~\ref{colcol}. The models depicted are those of Tsuji et
al. (\cite{tsuji04}), but in our analysis we also employed cloudless
models by Marley et al. (\cite{marley02}) and Burrows et
al. (\cite{burrows06}) obtaining similar results. The agreement
between the models and the field dwarf observations is reasonably
good. The great majority of the mid- and late-T dwarfs lie within the
log\,$g$\,=\,4.0 and 5.5~dex tracks, as expected for ``old'' dwarfs in
the solar neighborhood. This is also consistent with the recent
results of the spectral fitting work by Burgasser et
al. (\cite{burgasser06b}). These authors employed models by Burrows et
al. (\cite{burrows06}).

The near-infrared photometry of \sori~and current state-of-the-art
theory of ultracool dwarfs indicate that this object may have a
lower-gravity atmosphere than similarly classified T dwarfs in the
solar vicinity. Because of the different pressure and density
conditions at which H$_2$, CH$_4$, and CO absorptions are produced,
low-gravity objects tend to be brighter at $K$ and redder in all
near-infrared and [3.6]\,$-$\,[4.5] colors than comparable
high-gravity objects (see discussions and Figures in Knapp et
al. \cite{knapp04}; Patten et al. \cite{patten06}; Leggett et
al. \cite{leggett07}; Liebert \& Burgasser \cite{liebert07}). This is
what we qualitatively observe in \sori.

From the right panel of Fig.~\ref{colcol} and using the
solar-metallicity models by Tsuji et al. (\cite{tsuji04}), we derive
log\,$g$\,$\sim$\,3.0 dex and $T_{\rm
eff}$\,$\sim$\,1000--1100~K. This is consistent with previous results
obtained from the spectral fitting analysis of low- and
intermediate-resolution near-infrared spectra: $T_{\rm
eff}$\,$\sim$\,800$^{+200}_{-100}$ K and
log\,$g$\,$\sim$\,4.0\,$\pm$\,1.0 dex (Zapatero Osorio et
al. \cite{osorio02a}), $T_{\rm eff}$\,$\sim$\,1100$^{+200}_{-100}$ K
and log\,$g$\,$\sim$\,3.5\,$\pm$0.5 dex (Mart\'\i n \& Zapatero Osorio
\cite{martin03}), respectively. These authors compared observations to
theoretical data computed by Allard et
al. (\cite{allard01}). Recently, Liu et al. (\cite{liu07}) have
quantified the sensitivity of near-infrared spectra with $T_{\rm
eff}$\,$\sim$\,700--900~K to changes in metallicity and surface
gravity using a different grid of synthetic spectra by Burrows et
al. (\cite{burrows06}). \sori~is brighter at $K$ relative to $J$ or
$H$ by a factor of $\sim$1.4. Table~5 and Fig.~4 of Liu et
al. (\cite{liu07}) suggest that log\,$g$ is thus lower than the field
by about 1.0\,dex, in agreement with previous determinations. We
caution that current state-of-the-art synthetic spectra do not provide
detailed fits to the observed data (e.g., Burrows et
al. \cite{burrows06}). Therefore, any quantitative result derived from
the direct comparison of observations to models awaits further
confirmation. On the contrary, for a given temperature qualitative
predictions on the atmospheric relative behavior as a function of
gravity and metal content can be more reliable.

Metallicity might also be an issue. The \so~cluster has solar
abundance ([Fe/H]\,=\,0.0\,$\pm$\,0.1~dex; Caballero
\cite{caballero06}); we do not expect any metallicity effect when
comparing cluster members to the field. From theoretical
considerations, the effects of increasing abundance and decreasing
gravity on the near-infrared spectra of cool T dwarfs are
similar. From Fig.~20 of Burrows et al. (\cite{burrows06}), which
shows $(J-K)$ against $T_{\rm eff}$ for various gravities and metal
abundances, we infer a metallicity of [Fe/H]\,$\sim$\,$+$0.5 dex for
\sori~if its reddish effect was all due to metallicity. A similar rich
metal content is obtained from Liu et
al. (\cite{liu07}). Nevertheless, the super-solar metallicity
explanation, although possible, seems unlikely. On the one hand, the
metallicity distribution of F, G, and K dwarf stars in the solar
neighborhood peaks at around [Fe/H]\,=\,0.0 dex and roughly extends up
to $+$0.5~dex; less than $\sim$10\%~of the stars are more metal-rich
than $+$0.3 dex (Valenti et al. \cite{valenti05}; Santos et
al. \cite{santos05}; Boone et al. \cite{boone06}). On the other hand,
the IRAC/Spitzer data of the S\,Ori object do not support the high
metallicity case. Burgasser et al. (\cite{burgasser06b}) and Liebert
\& Burgasser (\cite{liebert07}) have demonstrated
2MASS\,J12373919$+$6526148 (T6.5) and 2MASS J09373487$+$2931409 (T6p)
to be old, high surface gravity brown dwarfs with sub-solar
abundance. These field dwarfs display [3.6]\,$-$\,[4.5] colors
slightly redder (by 0.15\,mag) than expected for their assigned
spectral types. In contrast to what could be inferred from the
near-infrared colors, this would indicate that \sori~is a low
metallicity T dwarf. Thus, a low-gravity atmosphere remains as the
most likely explanation to account for the observed photometry of
\sori.

\begin{figure*}
\centering
\includegraphics[width=9cm]{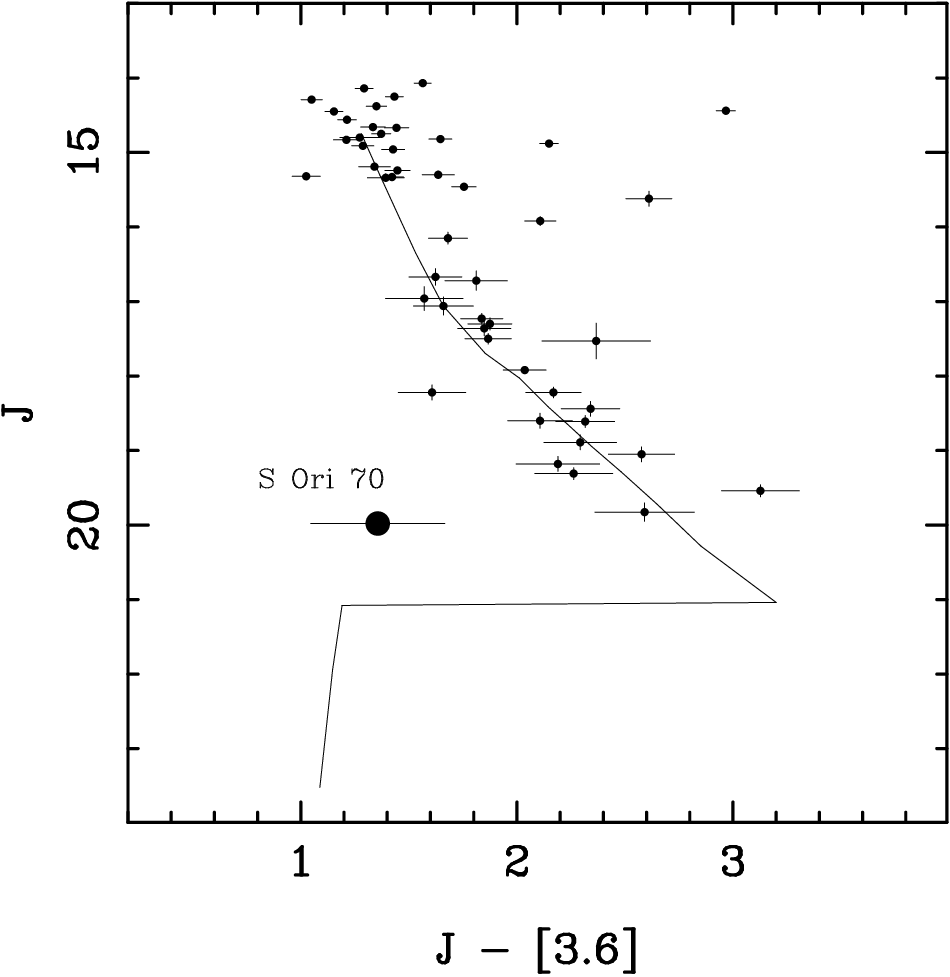}~~~
\includegraphics[width=9cm]{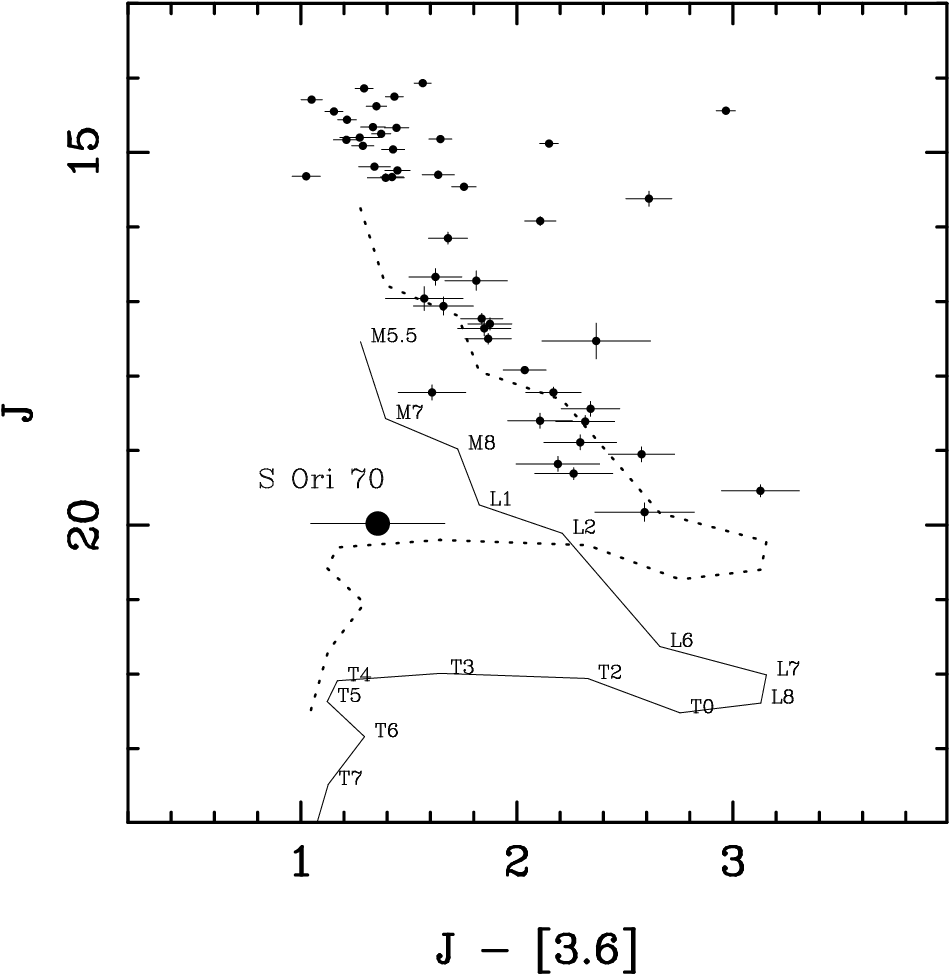}
\caption{Color-magnitude diagram of \so~low mass members. \sori~is
labeled. The 3-Myr isochrone by Chabrier \& Baraffe
(\cite{chabrier00}) is overplotted onto the data as a solid line in
the left panel (see text for $T_{\rm eff}$, log\,$L/L_\odot$
conversion into observables). The right panel shows the sequence of
M5.5--T8 field dwarfs at the distance of the cluster (solid line,
spectral types are indicated). The dotted line stands for the field
sequence shifted by $-$1.8 mag in the $J$-band to match the
photometric trend delineated by \so~members. Objects with
$J$\,$-$\,[3.6] colors significantly redder than the cluster sequence
show infrared flux excesses likely due to circum(sub)stellar disks
(Caballero et al. \cite{caballero07}).}
\label{colmag}
\end{figure*}

No obvious infrared flux excesses are detectable in the IRAC/Spitzer
[3.6] and [4.5] bands, suggesting that there is no envelope or disk
around \sori~emitting intensively at these wavelengths. A possitive
detection would have provided strong evidence for its youth. However,
we remark that disks around young, low-mass brown dwarfs (close to the
deuterium burning mass limit) are seen at wavelengths longer than
5\,$\mu$m (Luhman et al. \cite{luhman05}; Caballero et
al. \cite{caballero07}; Zapatero Osorio et al. \cite{osorio07}), while
the observed fluxes in the near-infrared up to 5\,$\mu$m are
photospheric in origin. The public [5.8]- and [8.0]-band IRAC/Spitzer
images are not conclusive for \sori.

\subsection{Color-magnitude diagram}

The photometric sequence of \so~substellar members, including \sori,
is shown in the $J$ vs $J$\,$-$\,[3.6] color-magnitude diagram of
Fig.~\ref{colmag} (Caballero et al. \cite{caballero07}; Zapatero
Osorio et al. \cite{osorio07}). This sequence follows a relatively
smooth progression with increasing color down to $J$\,$\sim$20 and
$J$\,$-$\,[3.6]\,=\,2.8~mag. The location of \sori~suggests that the
$J$\,$-$\,[3.6] index suddenly turns toward bluer values at a nearly
unchanged $J$ magnitude. A similar turnover (ocurring at spectral
types L7--L8, 1400--1300~K) is also observed in field ultracool
dwarfs. The field sequence of objects with spectral types M5.5--T8
moved to the distance of the \so~cluster is displayed in the right
panel of Fig.~\ref{colmag} (absolute magnitudes and colors are adopted
from Patten et al. \cite{patten06}, and references therein). Because
of their very young age, \so~low-mass stars and substellar objects are
in the phase of gravitational contraction (e.g., Chabrier \& Baraffe
\cite{chabrier00}). Cluster members thus show larger size and
luminosity than their older counterparts of related colors in the
field. As seen in Fig.~\ref{colmag}, the average cluster photometric
sequence appears brighter than the field by about 1.8~mag in the
$J$-band. The dotted line in Fig.~\ref{colmag} (right panel)
represents the field sequence normalized to the \so~locus of late-M
and early-L cluster members. \sori~nicely sits on the location
expected for \so~T-type members.

We have also compared our data to the {\sc cond} and {\sc dusty} solar
metallicity evolutionary models by Chabrier \& Baraffe
(\cite{chabrier00}). For the age range 1--8\,Myr, substellar objects
with $T_{\rm eff}$ between 700 and 1300~K, corresponding to the mass
interval $\sim$1--6\,\mj, show log\,$g$\,=\,3.0--4.0 dex, which
coincides within the large error bar with the surface gravity
estimation for \sori.

The 3-Myr isochrone (Chabrier \& Baraffe \cite{chabrier00}) is
displayed along the photometric sequence of \so~in the left panel of
Fig.~\ref{colmag}. Theoretical surface temperatures and luminosities
were converted into observed magnitudes and colors using the
color-temperature-spectral type and spectral type-bolometric
correction relationships given in the literature (Dahn et
al. \cite{dahn02}; Vrba et al. \cite{vrba04}; Knapp et
al. \cite{knapp04}; Patten et al. \cite{patten06}). The model
convincingly reproduces the cluster low mass sequence except for the
fact that \sori~appears overluminous by about 1~mag, which might
suggest binarity. This was also discussed in Zapatero Osorio et
al. (\cite{osorio02a}). However, there are issues that prevent us from
concluding whether this object is double or whether models make wrong
predictions for the smallest masses and young ages: {\sl (i)} as seen
from the field sequence (Fig.~\ref{colmag}, right panel), there is a
$J$-band brightening across the color turnover that theory fails to
reproduce (Vrba et al. \cite{vrba04}; Knapp et
al. \cite{knapp04}). {\sl (ii)} The blue color turnover takes place at
a roughly constant temperature in the field ($\sim$1300--1400~K), and
all color-temperature-bolometric correction transformations show a
sharp change at this point; on the contrary, the evolutionary models
available to us do not have a complete temperature sampling (this may
explain the abrupt color reversal at the bottom of the isochrone in
Fig.~\ref{colmag}). {\sl (iii)} The relations used to transform
theoretical predictions into observables are obtained for high-gravity
field objects. It is now known that gravity impacts significantly the
near- and mid-infrared colors of T dwarfs (Leggett et
al. \cite{leggett07}; Burrows et al. \cite{burrows06}), whereas the
colors of the warmer M and L types are not so sensitive to the gravity
parameter. Transformations are thus expected to be gravity-dependent
for the coolest temperatures. It becomes necessary to find a physical
explanation for the $J$ brightening feature and to discover more
\so~T-type, planetary-mass members for a proper comparison with
evolutionary tracks.

\section{Final remarks}
New near-infrared and Spitzer photometric data of
\sori~(T6\,$\pm$\,1.0) provide evidence that this object is young. Its
colors qualitatively follow the trend predicted for low gravity
atmospheres by state-of-the-art models. Its very low proper motion is
consistent with membership in the \so~cluster. The mass of \sori~is
then estimated from substellar evolutionary models to be within the
planetary regime in the range 2--7~\mj~(Zapatero Osorio et
al. \cite{osorio02a}; Mart\'\i n \& Zapatero Osorio
\cite{martin03}). This mass interval takes into account the object's
luminosity uncertainty and the most recent cluster age determination
(1--7~Myr; Sherry et al. \cite{sherry04}; Jeffries et
al. \cite{jeffries06}). The existence of such low-mass objects in
isolation at ages below 10~Myr is a challenging and interesting issue
for the theory of formation of stars, brown dwarfs and planets.

\begin{acknowledgements}
We are indebted to T. Tsuji and M. Marley for providing us
computer-ready files of their models. We also thank Sandy Leggett
(referee) for useful comments that helped us improve this work. Based
on observations collected at the Centro Astron\'omico Hispano Alem\'an
(CAHA) at Calar Alto, operated jointly by the Max-Planck Institut
f\"ur Astronomie and the Instituto de Astrof\'\i sica de Andaluc\'\i a
(CSIC). Some data were also obtained at the W. M. Keck Observatory,
which is operated as a scientific partnership between the California
Institute of Technology, the University of California, and NASA. The
Observatory was made possible by the generous financial support of the
W. M. Keck Foundation. We thank the Keck observing assistants and the
staff in Waimea for their kind support. The authors extend special
thanks to those of Hawaiian ancestry on whose sacred mountain we are
privileged to be guests. This research has made use of data products
from the Two Micron All Sky Survey, which is a joint project of the
University of Massachusetts and the Infrared Processing and Analysis
Center/California Institute of Technology, funded by the National
Aeronautics and Space Administration and the National Science
Foundation. This work is based in part on archival data obtained with
the Spitzer Space Telescope, which is operated by the Jet Propulsion
Laboratory, California Institute of Technology under a contract with
NASA. Support for this work was provided by an award issued by
JPL/Caltech. Partial financial support was provided by the Spanish
projects AYA2003-05355 and AYA2006-12612.
\end{acknowledgements}

\newpage

\begin{appendix} 

\section {Filter passbands}

\begin{figure*}
\centering
\includegraphics[width=5.9cm]{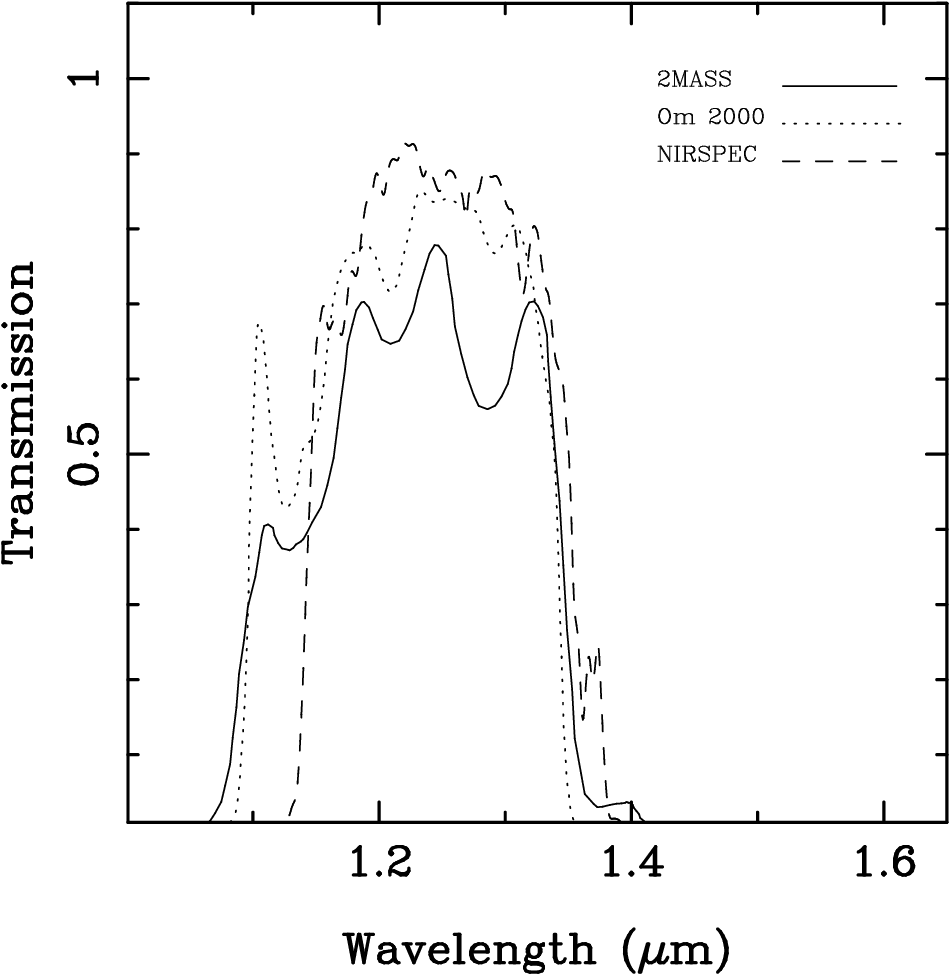}~~
\includegraphics[width=5.9cm]{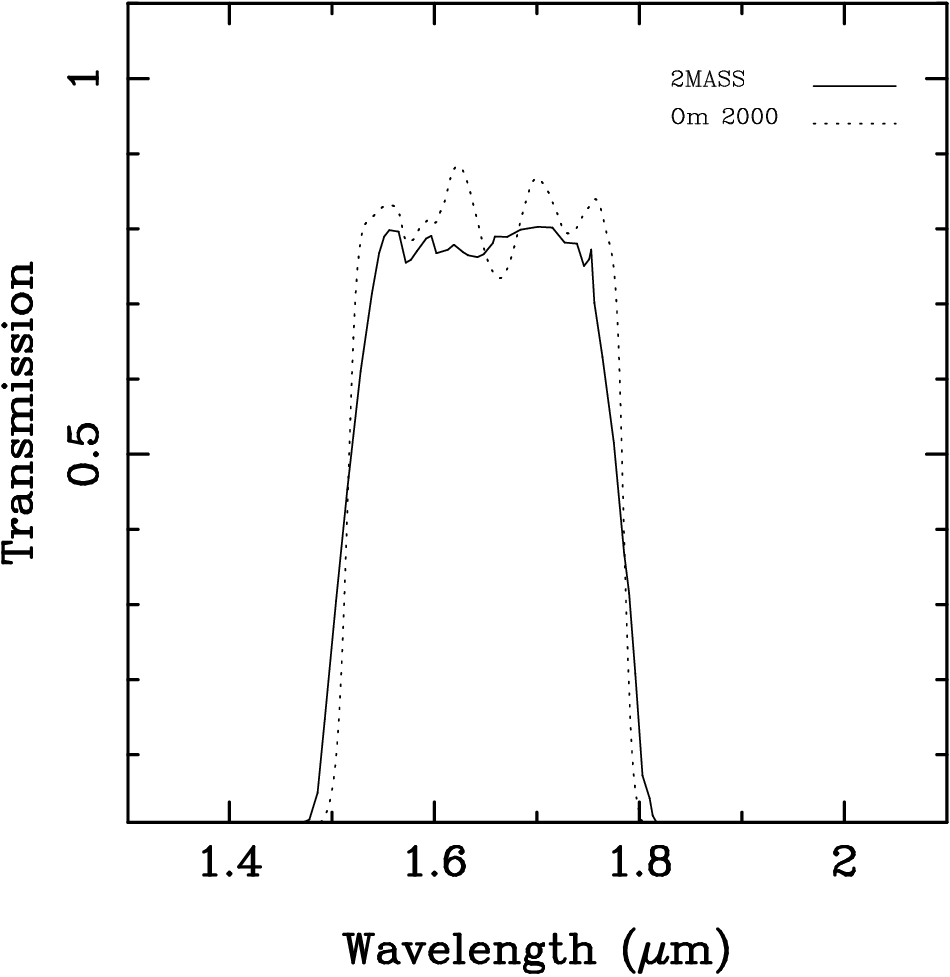}~~
\includegraphics[width=5.9cm]{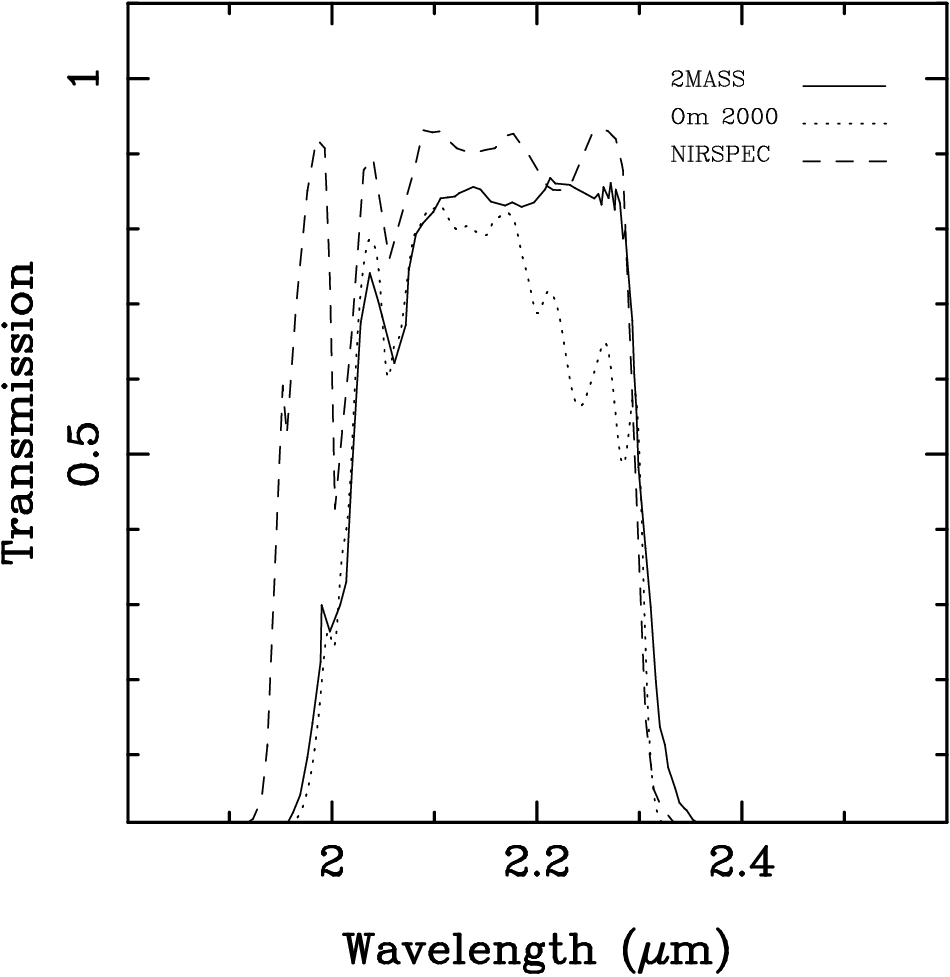}
\caption{Passbands of the $J$ (left), $H$ (middle), and $K$ (right) filters used in this work convolved with the observatories atmospheric transmission curves. Labels correspond to the NIRSPEC (Keck) and Omega-2000 (CAHA 3.5-m) instruments, and 2MASS filters. Data have been taken from the observatories web pages. The Omega-2000 filter passbands were convolved with the 2MASS atmospheric transmission curve because the latter information is not available on the CAHA web site.}
\label{filters}
\end{figure*}

\end{appendix}


\begin{thebibliography}{}

\bibitem[2001]{allard01}
  Allard, F., Hauschildt, P. H., Alexander, D. R., Tamanai, A., \& 
  Schweitzer, A. 2001, ApJ, 556, 357
\bibitem[2006]{allers06}
  Allers, K. N., Kessler-Silacci, J. E., Cieza, L. A., \& Jaffe, D. T.
  2006, ApJ, 644, 364
\bibitem[2006]{artigau06}
  Artigau, E., Doyon, R., Lafreni\`ere, D., Nadeau, D., Robert, J., 
  \& Albert, L. 2006, ApJ, 651, L57
\bibitem[2000]{bailer00}
  Bailer-Jones, C. A. L., Bizenberger, P., \& Storz, C. 2000, Proc. 
  SPIE, 4008, 1305
\bibitem[1998]{baraffe98}
  Baraffe, I., Chabrier, G., Allard, F., \& Hauschildt, P. H. 1998,
  A\&A, 337, 403
\bibitem[2006]{boone06}
  Boone, R. H., King, J. R., \& Soderblom, D. R. 2006, New A. Rev., 50, 526
\bibitem[2004]{burgasser04}
  Burgasser, A. J., Kirkpatrick, J. A., McGovern, M. R., McLean, I. S., 
  Prato, L.,  \& Reid, I. N. 2004, ApJ, 604, 827
\bibitem[2006a]{burgasser06a}
  Burgasser, A. J., Geballe, T. R., Leggett, S. K., Kirkpatrick, J. D., \& 
  Golimowski, D. A. 2006a, ApJ, 637, 1067
\bibitem[2006b]{burgasser06b}
  Burgasser, A. J., Burrows, A., \& Kirkpatrick, J. D. 2006b, ApJ, 639, 1095
\bibitem[1997]{burrows97}
  Burrows, A., Marley, M., Hubbard, W. B., Lunine, J. L., Guillot, T., 
  Saumon, D., Freedman, R., Sudarsky, D., \& Sharp, C. 1997, ApJ, 491, 856
\bibitem[2006]{burrows06}
  Burrows, A., Sudarsky, D., \& Hubeny, I. 2006, ApJ, 640, 1063
\bibitem[2006]{caballero06}
  Caballero, J. A. 2006, Ph. D. Thesis, Univ. La Laguna (Tenerife, Spain)
\bibitem[2007]{caballero07}
  Caballero, J. A., et al. 2007, A\&A, 470, 903
\bibitem[2000]{chabrier00}
  Chabrier, G., \& Baraffe, I. 2000, ARA\&A, 38, 337
\bibitem[2004]{chauvin04}
  Chauvin, G.,  Lagrange, A.-M, Dumas,C., Zuckerman, B., Mouillet, D., 
  Song, I., Beuzit, J.-L., \&  Lowrance, P. 2004, A\&A, 425, L29
\bibitem[2005]{chauvin05}
  Chauvin, G., Lagrange, A.-M, Dumas,C., Zuckerman, B., Mouillet, D., 
  Song, I., Beuzit, J.-L., \&  Lowrance, P. 2005, A\&A, 438, L25
\bibitem[2007]{covey07}
  Covey, K. R., Ivezi\'c, Z., Schlegel, D., et al. 2007, ApJ, in press
\bibitem[2002]{dahn02}
  Dahn, C. C., et al. 2002, AJ, 124, 1170
\bibitem[1999]{zeeuw99}
  de Zeeuw, P. T., Hoogerwerf, R., de Bruijne, J. H. J., Brown, A. G. A., \& 
  Blaauw, A. 1999, AJ, 117, 354
\bibitem[2004]{fazio04}
  Fazio G. G., Hora, J. L., Allen, L. E., et al. 2004, ApJ Suppl., 154, 10
\bibitem[2002]{geballe02}
  Geballe, T. R., Knapp, G. R., Leggett, S. K., et al. 2002, ApJ, 564, 466
\bibitem[2006]{gonzalez06}
  Gonz\'alez-Garc\'\i a, B. M., Zapatero Osorio, M. R., B\'ejar, 
  V. J. S., Bihain, G., Barrado y Navasdcu\'es, D., Caballero, J. A., 
  \& Morales-Calder\'on, M. 2006, A\&A, 460, 799
\bibitem[2007]{hernandez07}
  Hern\'andez, J., Hartmann, L., Megeath, T., et al. 2007, ApJ, 662, 1067
\bibitem[2006]{ray06} 
  Jayawardhana, R., \& Ivanov, V. D. 2006, Science, 313, 1279
\bibitem[2006]{jeffries06}
  Jeffries, R. D., Maxted, P. F. L., Oliveira, J. M., \& Naylor, T. 
  2006, MNRAS, 371, L6
\bibitem[2004]{knapp04}
  Knapp, G., et al. 2004, AJ, 127, 3553
\bibitem[1999]{leggett99}
  Leggett, S. K., Toomey, D. W., Geballe, T. R., \& Brown, R. H. 1999, ApJ, 
  517, L139
\bibitem[2002]{leggett02}
  Leggett, S. K., et al. 2002, ApJ, 564, 452
\bibitem[2007]{leggett07}
  Leggett, S. K., Saumon, D., Marley, M. S., Geballe, T. R., \& Fan, X. 
  2007, ApJ, 655, 1079
\bibitem[2007]{liebert07}
  Liebert, J., \& Burgasser, A. J. 2007, Apj, 655, 522
\bibitem[2006]{liu06}
  Liu, M. C., Leggett, S. K., Golimowski, D. A., Chiu, K., Fan, X., Geballe, 
  T. R., Schneider, D. P., \& Brinkmann, J. 2006, ApJ, 647, 1393
\bibitem[2007]{liu07}
  Liu, M. C., Leggett, S. K., \& Chiu, K. 2007, ApJ, 1507
\bibitem[2007]{looper07}
  Looper, D. L., Kirkpatrick, J. D., \& Burgasser, A. J. 2007, AJ, 134, 1162
\bibitem[2000]{lucas00}
  Lucas, P. W., \& Roche, P. F. 2000, MNRAS, 314, 858
\bibitem[2005]{lucas05}
  Lucas, P. W., Roche, P. F., \& Tamura, M. 2005, MNRAS, 361, 211
\bibitem[2005]{luhman05}
  Luhman, K. L., Adame, L., D'Alessio, P., Calvet, N., Hartmann, L., 
  Megeath, S. T., \& Fazio, G. G. 2005, ApJ, 635, L93
\bibitem[2007]{luhman07}
  Luhman, K. L., et al. 2007, ApJ, 654, 570
\bibitem[2002]{marley02}
  Marley, M. S., Seager, S., Saumon, D., Lodders, K., Ackerman, A. S., 
  Freedman, R., \& Fan, X. 2002, ApJ, 568, 335
\bibitem[2000]{mclean00}
  McLean, I. S., et al. 2000, ApJ, 533, L45
\bibitem[2003]{martin03}
  Mart\'\i n, E. L., \& Zapatero Osorio, M. R. 2003, ApJ, 593, L113
\bibitem[2006]{mugrauer06}
  Mugrauer, M., Seifahrt, A., Neuh\"auser, R., \& Mazeh T. 2006, MNRAS, 
  373, L31
\bibitem[2002]{oliveira02}
  Oliveira, J. M., Jeffries, R. D., Kenyon, M. J., Thompson, S. A., \& Naylor, 
  T. 2002, A\&A, 382, L22
\bibitem[2006]{patten06}
  Patten, B. M., et al. 2006, ApJ, 651, 502
\bibitem[1997]{perryman97}
  Perryman, M. A. C., et al. 1997, A\&A, 323, L49
\bibitem[1998]{pickles98}
  Pickles, A. J. 1998, PASP, 110, 863
\bibitem[2005]{santos05}
  Santos, N. C., Israelian, G., Mayor, M., Bento, J. P., Almeida, P. C., 
  Sousa, S. G., Ecuvillon, A. 2005, A\&A, 437, 1127
\bibitem[2004]{sherry04}
  Sherry, W. H., Walter, F. M., \& Wolk, S. J. 2004, AJ, 128, 2316
\bibitem[2004]{stephens04}
  Stephens, D. C., \& Leggett, S. K. 2004, PASP, 116, 9
\bibitem[2005]{tinney05}
  Tinney, C. G., Burgasser, A. J., Kirkpatrick, J. D., \& McElwain, M. W. 
  2005, AJ, 130, 2326
\bibitem[2004]{tsuji04} 
  Tsuji, T., Nakajima, T., \& Yanagisawa, K. 2004, ApJ, 607, 511
\bibitem[2005]{valenti05}
  Valenti, J. A., \& Fischer, D. A. 2005, ApJ Suppl., 159, 141
\bibitem[2004]{vrba04}
  Vrba, F. J., et al. 2004, AJ, 127, 2948
\bibitem[2000]{osorio00}
  Zapatero Osorio, M. R., B\'ejar, V. J. S., Mart\'\i n, E. L., Rebolo, 
  R., Barrado y Navascu\'es, D., Bailer-Jones, C. A. L., \& Mundt, R. 2000, 
  Science, 290, 103
\bibitem[2002a]{osorio02a}
  Zapatero Osorio, M. R., B\'ejar, V. J. S., Mart\'\i n, E. L., Rebolo, 
  R., Barrado y Navascu\'es, D., Mundt, R., Eisl\"offel, J., \& Caballero, 
  J. A. 2002a, ApJ, 578, 536 (Paper~I)
\bibitem[2002b]{osorio02b}
  Zapatero Osorio, M. R., B\'ejar, V. J. S., Pavlenko, Ya., Rebolo, R., 
  Allende Prieto, C., Mart\'\i n, E. L., \& Garc\'\i a L\'opez, R. 2002b, 
  A\&A, 384, 937
\bibitem[2007]{osorio07}
  Zapatero Osorio, M. R., Caballero, J. A., B\'ejar, V. J. S., et al. 
  2007, A\&A, 472, L9
\end{thebibliography}
\end{document}